\documentstyle[prd,aps,floats,times,epsfig,12pt]{revtex}
\def\be{\begin{equation}}
\def\ee{\end{equation}}
\def\bea{\begin{eqnarray}}
\def\eea{\end{eqnarray}}
\begin{document}
\title{Quintessence Model With Double Exponential Potential}
\author{A.A.Sen{\footnote{anjan@mri.ernet.in} and S.Sethi\footnote{sethi@mri.ernet.in}}}
\address{Harish-Chandra Research Institute,Chhatnag Road, Jhusi, Allahabd 211019, India}
\date{\today}
\maketitle
\begin{abstract}
We have reinvestigated the quintessence model with minimally coupled scalar
field in the context of recent Supernova observation at $z=1.7$. By assuming
the form of the scale factor which gives both the early time deceleration and
late time acceleration, consistent with the observations, we show that
one needs a double exponential potential. We have also shown that the equation
of state and the behaviour of dark energy density are reasonably consistent
with earlier constraints obtained by different authors. This work shows again
the importance of double exponential potential for a quintessence field.
\end{abstract}
\vspace{10mm}
Over the first few years we are experiencing some of the most interesting
cosmological observations. Data from the luminosity distance-redshift
observations of the type Ia Supernova (SNIa) collected by two survey teams
,The Supernova Cosmology Project and the High-z Supernova Search team,
\cite{super1,super2} predict that the universe is currently going through an
accelerating expansion phase. Although there are two different interpretation
of the Supernova observations ---- intergalactic dust and SN luminosity
evolution \cite{dust}, the recent observations of SN 1997ff, at $z \sim 1.7$
\cite{newriess} put the accelerating Universe hypothesis on a firm footing. 
This also provides the first evidence for an early epoch of decelerating
universe. On the the hand, the recent observations of the acoustic peaks of
the Cosmic Microwave Background (CMB) temperature fluctuations
\cite{cmb}favour a spatially flat universe, as predicted by the inflationary
models. 

If the results of these two observations are put together, one
immediate conclusion is that the energy density of the universe is currently
dominated by a  form of matter having negative pressure, commonly referred
as "{\it dark energy}". This component is qualitatively different from the
standard dark matter in the sense that it has large negative pressure and it
is approximately homogeneous, not clustering with matter on scales of clusters
of galaxies. The first and obvious choice for this dark energy component is
the cosmological constant $\Lambda$ which represents the energy of a quantum
vacuum. However the problem of $\Lambda$ being the dominant component of the
total energy density stems from the fact that the energy scale involved is
lower than the normal energy scale of most particle physics model by a factor
$\sim 10^{-120}$.

So to find some alternative candidate for this acceleration a dynamical 
$\Lambda$\cite{cal} in the form of a scalar field with some self interacting 
potential\cite{peebetc} is considered whose slowly varying energy density
mimics an effective cosmological constant. The idea of this  candidate,called
{\it quintessence}\cite{cal}, is borrowed from the inflationary phase of the
early universe, with the difference that it evolves at a much lower energy
scale. The energy density of this field, though dominant at present epoch,
must remain subdominant at very early stages and has to evolve in such a way
that it becomes comparable with the matter density $\Omega_m$ now. This type
of specific evolution, needs several constraints on the initial conditions 
and fine tuning of parameters for the potential. A new form 
of quintessence field called ``{\it tracker field}''\cite{zla} has been 
proposed to solve this problem. It has an equation
 of motion with an attractor like solution in a sense that for a 
wide range of initial conditions the equation of motion converges 
to the same solution. 

There are a number of quintessence models which have been put forward in
recent years. They involve a scalar field rolling down its potential 
\cite{wett}-\cite{rubano}, an axion field \cite{axion}, scalar tensor theories
of gravity \cite{bertol}-\cite{sen4}, dilaton in context of string theory
\cite{dil}, and also fields arising from compactifications of the
multidimensional Einstein-Yang-Mills system \cite{eym}. 
In a very recent work, Zimdahl and Pavon\cite{zipav} have shown that a
suitable coupling between a minimally coupled quintessence field and the
pressureless cold dark matter gives  a constant ratio of the energy densities
of both components which is compatible with the late time acceleration of the
universe. They have termed it "{\it interacting quintessence}". In another
recent work, Tocchini-Valentini and Amendola have investigated the cosmological
models when this coupling  between the quintessence field and the perfect
fluid dark matter is linear \cite{tochini}.

Although all of these have their own merits in explaining the dark energy of
the universe, there are number of difficulties with these models. One of them
is to smoothly match the current accelerating universe with matter or
radiation dominated decelerated universe. The current accelerated expansion is
obviously a recent phenomena as one needs a sufficiently long matter dominated
decelerated phase which should last until a recent past for the observed
structure to develop from the density inhomogeneities. Further the success of
big bang nucleosynthesis gives us a strong evidence of the radiation dominated
decelerated phase when the universe is few seconds old. Although this required
feature of the decelerated expansion is by far observationally untested, the
recent observation of the SN1997ff at $z=1.7$ \cite{newriess} confirms this
essential feature of the history of the universe. In a recent analysis,
using a new technique which is independent of the content of the universe,
Turner and Riess \cite{turriess} have shown that supernova data favour past
deceleration ($z>0.5$) and a recent acceleration ($z<0.5$).

Also there are a number of investigations in order to constrain the equation
of state of the dark energy component taking both the supernova data and data
for the cmb measurements into account \cite{eqn}. In one of the recent
analysis, Corassaniti and Copeland \cite{copecross} have shown that most of the potentials used
so far, including the inverse power law one as well as the Supergravity
inspired potential, are not satisfactory as far as these constraints on the
equation of state are concerned.

In this work, we have investigated these issues of the dark energy in a
different way. We have taken the dark energy to be a minimally coupled scalar
field rolling down its potential. Instead of assuming the form of its
potential, we  have assumed the form of the scale factor ( which in
turn gives the form of the Hubble parameter) keeping in mind that
although the universe is presently accelerating but it was decelerating in
recent past. In their work, Turner and Riess have emphasised \cite{turriess}
the importance of assumption about $H(z)$ in order to use the SNe data to
probe the history of the universe. 
This method of finding exact solutions for scalar field cosmology was first used by Ellis and Madsen \cite{ellis} for inflationary models. They showed that one can determine the potential which gives the best behaviour in terms of its implications for cosmology. Later Uggla et.al have discussed this method for a more generalised situation \cite{uggla}
Assuming some specific form of the scale
factor which gives both the decelerating universe in the past as well as the
accelerating one at present, we have tried to fit our model with the SNe data
including the recent data at $z=1.7$.  For particular value of the
parameter in our model for which our model fits reasonably well with the
observational data, we have found that the potential turns out to be a double
exponential one. This sort of potential has been considered earlier by
different authors for quintessence models \cite{barr,rubano}. We have also 
shown that energy density for the scalar field remains  sufficiently below that of the
matter field for higher redshifts  and starts dominating in the recent
past.

Let us consider a spatially flat, homogeneous, isotropic universe, with a
pressureless (dust) matter fluid and a scalar field $\phi$ with potential
$V(\phi)$ minimally coupled with gravity. The equations of motion are given by

\bea
3H^{2} = 8\pi G[\rho_{m}+{1\over{2}}\dot{\phi^{2}}+V(\phi)]\\
\ddot{\phi}+3H\dot{\phi}+V^{'}(\phi) = 0\\
\dot{\rho_{m}}+3H\rho_{m} = 0
\eea
where $\rho_{m}$ is the energy density for the matter fluid and
$H=\dot R(t)/R(t)$ is the Hubble parameter and  $R(t)$ is the scale
factor. Here overdot and prime mean differentiations with respect to time and
scalar field $\phi$ respectively. In this system of equations, one has three
independent equations and four unknowns, which demands one assumption about
the unknowns to solve the system. In most of the previous works with minimally
coupled scalar field, the form of the potential has been assumed in order to
solve the system. Although the assumption of the form of the potential from
the particle physics viewpoint is a reasonable way, many a times it leads to
complicated equations for the scale factor to solve and also to study
different observable quantities. In this work, we proceed in a different
way. The Supernova observations, made by two teams (Perlmutter et.al and Riess
et.al), favour a present day accelerating universe. Also the recent
observation of SN 1997ff at $z=1.7$ provides the evidence of decelerating
universe around that redshift. In a recent paper, Turner and Riess have shown
that SN data favour a recent acceleration ($z<0.5$) and a past deceleration
($z>0.5$). Keeping this in mind, we assume scale factor $R(t)$ of the form

\be
R(t) = {R_{0}\over{\alpha}}\left[\sinh(t/t_{0})\right]^{\beta}
\ee

where $t_{0}$ is the present time, $R_{0}=R(t=t_{0})$ is the present day
scale factor and $\alpha=\left[\sinh(1)\right]^{\beta}$, $\beta$ being a
constant. The interesting feature of this scalar factor is that for $\beta<1$,
the universe is decelerating for $t<<t_{0}$, and exponentially accelerating
for $t>>t_{0}$. As an example, for $\beta=2/3$, 
\bea
R(t) \propto t^{2/3} \hspace{2mm} \hbox{for} \hspace{2mm} t<<t_{0}\\
R(t) \propto \exp(t) \hspace{2mm} \hbox{for} \hspace{2mm} t>>t_{0}
\eea

In terms of redshift, the expression for the Hubble parameter is given by 

\be
H(z) = {\beta\over{t_{0}}}\left[1+(y/\alpha)^{2/\beta}\right]^{1/2}\
\ee

where $y=1+z$. One can also calculate the deceleration parameter $q(z)$ in our
model which is given by

\be
q(z) = {1\over{\beta}}\left[{(y/\alpha)^{2/\beta}\over{(y/\alpha)^{2/\beta}+1}}\right] - 1
\ee
\begin{figure}[h]
\centering
\leavevmode\epsfysize=7cm \epsfbox{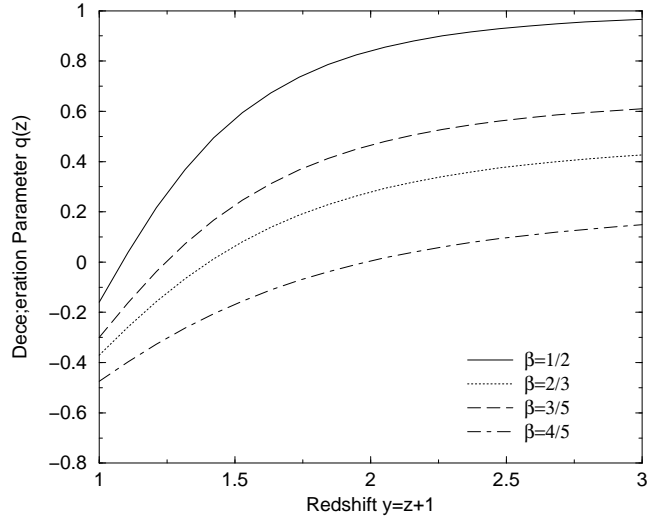}
\caption{The deceleration parameter $q$ {\it vs} redshift $z$}
\label{fig1}
\end{figure}
In fig 1  we have plotted the deceleration parameter q(z) for different values of $\beta$. The
figure shows that although the universe is accelerating at present, it was
decelerating in recent past. For all the different values of $\beta$ we have
used in the figure, the universe is decelerating at $z=1.7$ which is in
agreement with the recent observation. 

Astronomers measure luminosities in logarithmic units, called magnitudes
defined by
\be
{m_B}(z) = {\cal{M}} + 5\log_{10}(D_{l})
\ee
where ${\cal{M}}= M - 5\log_{10}(H_{0})$ and $ D_{l} = H_{0}d_{l}$ with $M$ is
the absolute luminosity of the object and $d_{l}$ is the luminosity distance
defined by
\be
d_{l} = R(t_{0})(1+z)r_{1}
\ee 
for an event at $r=r_{1}$ and at time $t=t_{1}$.  One can show that for nearby
sources (in the low redshift limit) the equation (9) can be written as
\be
{m_B}(z) =  {\cal{M}} + 5\log_{10}z
\ee
which can be used to measure the ${\cal{M}}$ by using low-redshifts
supernovae-measurements. In fig 2 we have plotted the ${m_B}(z)$ for different
values of $\beta$.

\begin{figure}[h]
\centering
\leavevmode\epsfysize=7cm \epsfbox{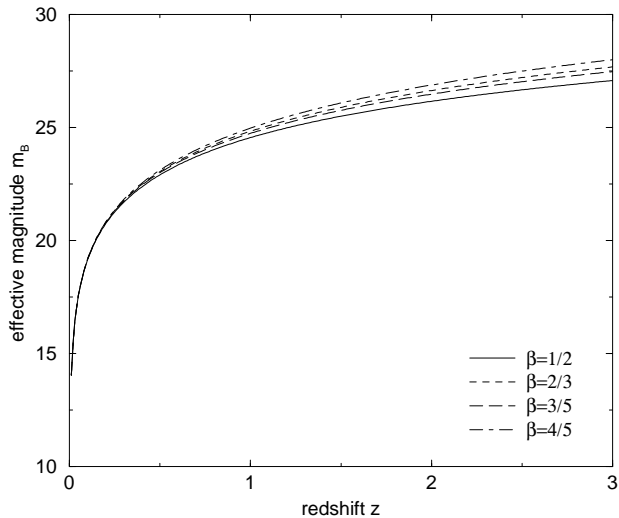}
\caption{The effective magnitude $m_{B}$ {\it vs} redshift $z$}
\label{fig2}
\end{figure}

We now obtain the best-fit value of $\beta$ by comparing our model
predictions with the SN1a data. We use the high-z data of the
Supernova Cosmology Project (SCP) by Permutter et al. \cite{super1}  
and the low-z data from Calan-Tolado survey \cite{hamuy} 
for our study. Of the 60 data point points, we use 
54 data points for our analysis (Fit C--D  of the SCP data;
for details of the excluded data points see Perlmutter {\it et al.}
1998). In addition we use the $z \simeq 1.7$ datum reported by
Riess \cite{newriess} in 2001. 
The best-fit value and 1$\sigma$ errors  from the SN1a data are $\beta =
0.81^{{+}0.18}_{{-}0.16}$; the best-fit   
$\chi^2/{\it dof} = 1.08$.  In this analysis we marginalize over
$H_0$.

In Fig~3, we show the $\chi^2/{\it dof}$
as a function of the parameter $\beta$. 
But as the parameter $\beta$ effectively determines the turnover point from acceleration to  deceleration, given in equation (8) by $q(z)=0$, one can interpret this as a likelihood analysis of the turnover redshift.
Fig~3  shows that the value of
$\chi^2$ is not very sensitive to the value of $\beta$ for $0.5 \simeq
\beta \simeq 1$. 
This can also be seen in Fig 2 where we have plotted the effective magnitude $m_{B}$ with respect to the redshift $z$ for different choice of $\beta$. There also,  it is very difficult to distinguish models with different values of $\beta$ upto redshift $z \sim 1$.
This means, from Fig~1, that the epoch at which
the universe passes from the accelerating to the decelerating phase is
not very well determined. We can compare our conclusions with the
results of Turner and Riess \cite{turriess} who claim that the
universe is accelerating for $z \le 0.5$ and is decelerating for higher
redshifts. Though their result is consistent with our conclusions
($\beta \simeq  0.65$ implies  (Fig~1) the results of Turner and Riess
\cite{turriess} and it is within 1$\sigma$ of our best fit value), we
cannot conclude it. On the other hand, our results are in greater
accord their assertion that the deceleration parameter is increasing
as the redshift increases; it is evident from the range of allowed
$\beta$ values in our analysis (Fig~1).

\begin{figure}[h]
\centering
\epsfysize=7cm \epsfbox{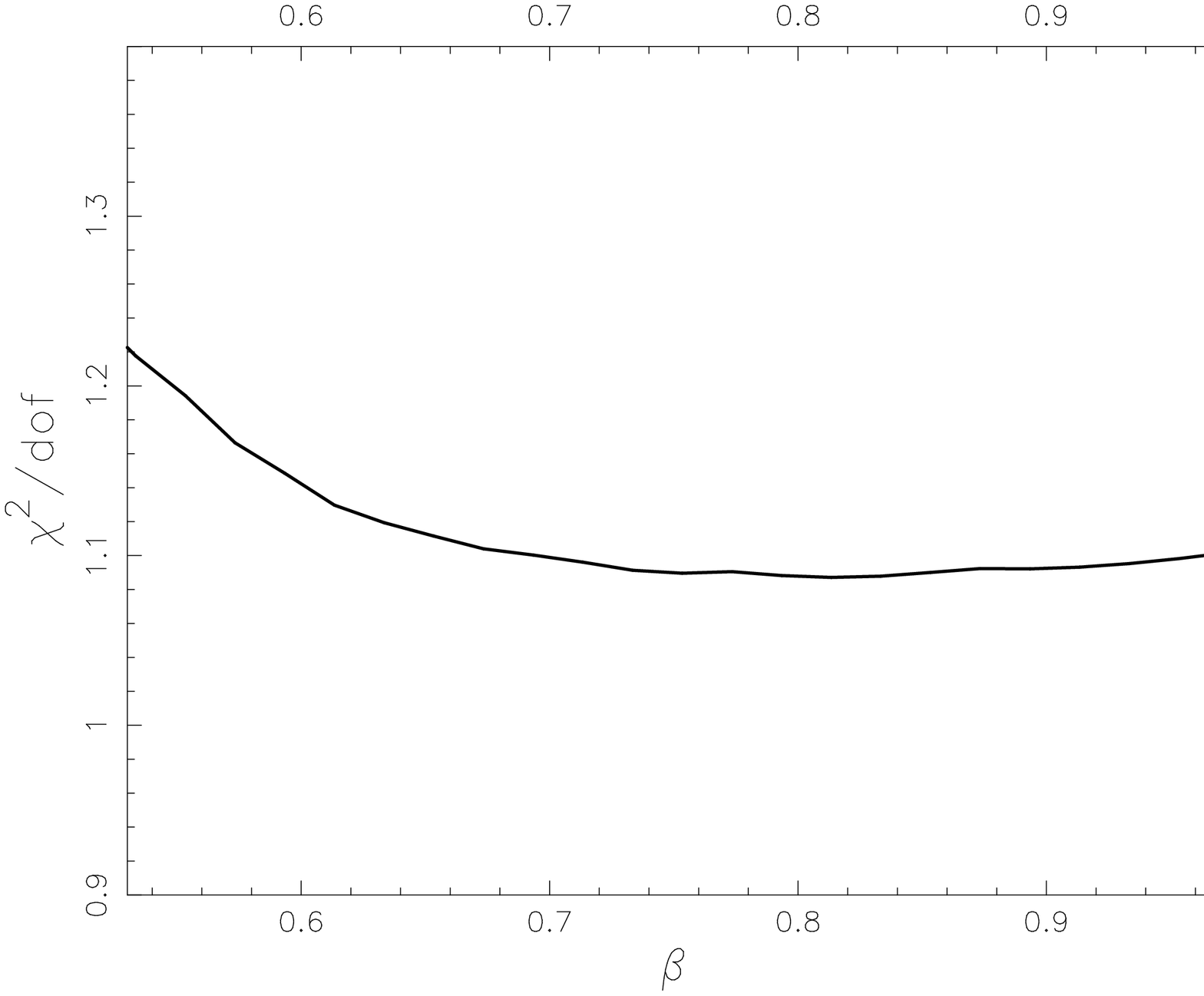}
\caption{$\chi^2/{\it dof}$ vs $\beta$}
\label{fig3}
\end{figure}

In this paper,  we use only $\beta = 2/3$; this value is
within $1\sigma$ of the best fit value. It is important to use this value of $\beta$ in order to have a early time matter dominated decelerated universe. The age of the universe with this choice of parameter turns out to be approximately 14 GYrs with $H_{0}=0.6\times 10^{-10}$ per yr.

Now using equations (1)-(4) one can write
\be
2\dot{H}+3H^{2} = 8\pi G[- {1\over{2}}\dot{\phi^{2}}+V(\phi)]={4\over{3t_{0}^{2}}} \nonumber
\ee
which gives
\be
V(\phi) = {1\over{2}}\dot{\phi}^{2}+{1\over{6\pi Gt_{0}^{2}}}
\ee
From the above equation, one can write
\be
V^{'}(\phi) = \ddot{\phi}
\ee
Using this relation in the scalar field wave equation (2) one can solve $\phi$
as
\be
\phi = At_{0}\log_e\left[\tanh(t/2t_{0})\right]
\ee
where $A$ is a constant of integration. Using the expression $\phi$ and
equation (13) one can get the form of the potential which is given by

\be
V(\phi) = {A^{2}\over{8}}\left(e^{2a\phi}+e^{-2a\phi}\right) + V_{0}
\ee

where $a={1\over{At_{0}}}$ and $V_{0}={1\over{6\pi Gt_{0}^{2}}}-{A^2\over{4}}$.
This type of potential has been earlier used by Barreiro et al.\cite{barr} and Rubano
et al.\cite{rubano}. In a recent paper \cite{copecross}, Corasaniti and Copeland have used the Supernova data
(excluding the recent data at $z=1.7$) and measurements of the position of the
acoustic peaks of the CMBR spectra to constrain a general class of potentials
including the inverse power law models and the recently proposed Supergravity
inspired potential. They have argued that in order to have the equation of state parameter $\omega_{\phi} \sim -1$, the quintessence field has to undergo damped oscillations around the minimum of the potential or has to evolve in a very flat region of the potential. And in this respect double exponential potential can be a good choice although they have not included it in their analysis.
 
The expression for the energy density $\rho_{\phi}$ for the scalar field and
the equation of state $\omega_{\phi}$ for the scalar field are given by
\bea
\rho_{\phi} &=& A^{2}\sinh^{-2}(t/t_{0})+{1\over{6\pi Gt_{0}^{2}}}\\
\omega_{\phi} &=& -{1\over{6\pi Gt_{0}^{2}A^{2}\sinh^{-2}(t/t_{0})+1}}
\eea

Using the expression for $\rho_{\phi}$, $H$ and $\rho_{m}=\rho_{m0}(1+z)^{3}$ in equation (1), one can get

\be
A^{2}={1\over{6\pi Gt_{0}^{2}}}-\rho_{m0}\alpha^{3}
\ee
Also if one has to the link the potential given in (16) with that of Rubano et.al\cite{rubano} one has to set $A^2 t_{0}^2={1\over{3\pi G}}$ which is not possible from equation (19). 

\begin{figure}[h]
\centering
\leavevmode\epsfysize=7cm \epsfbox{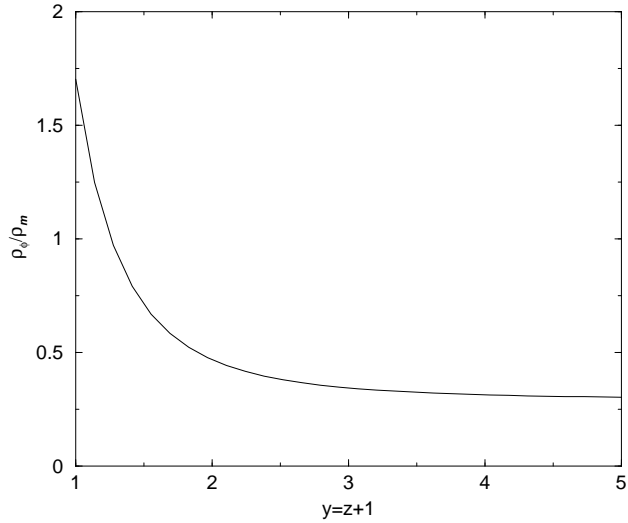}
\caption{The ratio of the energy densities  {\it vs} redshift $z$}
\label{fig4}
\end{figure}

Baccigalupi et.al \cite{bacci2} have shown in recent work that the position of the first doppler peak prefers a quintessence model with $\omega_{\phi0}\sim -0.8$ for the prior $\Omega_{\phi}=0.7$. If we use this value of $\Omega_{\phi}$ in our model 
the constraint on the constant $A$ turns out to be  $6\pi Gt_{0}^{2}A^{2} = 0.286$ and  the equation of state
for the scalar field at present $\omega_{\phi}(z=0)$ comes out to be $\sim
-0.83$ which is reasonably consistent with the earlier bound on
$\omega_{\phi}(z=0)$ obtained by different authors\cite{bacci2,eqn}. 

In fig 4 we have
plotted the ratio of the two energy densities $\rho_{\phi}/\rho_{m}$. One can look at this figure to see that the ratio is approximately constant and also much less than one for the higher redshift. It starts dominating in the recent time (around $z=0.5$). It shows that the scalar field scales below the matter energy density in the early universe. This is feature is quite similar to that of the tracker field discussed earlier.

In conclusion, we have re-investigated the role of double exponential potential
for the quintessence field in the context of the recent observation of
SN1997ff at $z=1.7$ but in a different manner. We have assumed the form of the scale 
factor for which  the universe interpolates between the early time decelerated
expansion and the late time accelerated expansion. We then tried to fit our
model with the SN1a observation including the recent observation at $z=1.7$. 
There are two free parameters in our model: $H_{0}$ and $\beta$. We have marginalisd over $H_{0}$ in our analysis. We then 
have obtained that the best fit value of the parameter $\beta$ appearing in
the form of the scale factor. 
Constancy of $\chi^2$ over a wide values of $\beta$ does suggest that
it is difficult to pinpoint the exact turn around from acceleration
to deceleration.  It is largely owing the quality of the data.
It is not possible  to say if the universe is decelerating or accelerating
taking points up to z = 0.83 (Milne universe with $a(t) \propto t$ is
not such a bad fit to the data up to z = 0.83, as shown in the
original paper of Perlmutter et al.).
However inclusion of the new data point at
z = 1.7 suggests that the  deceleration
parameter is increasing  as the  redshift increases  (this also happens to be
the strongest claim of Turner and Riess). Given the insensitivity of
$\chi^2$ on $\beta$, this  is also one of our main conclusion of our analysis.

To have a early matter dominated decelerated universe  we have
chosen  $\beta=2/3$ for which our model fits reasonably with
observation. We have then showed that the potential one needs is a double
exponential potential. We have also investigated the different relevant
parameters, such as the equation of state and the ratio of the two energy
densities and showed that they behave reasonably well as far as the
consistency of the model is concerned. Although this method of solving the field equations is completely {\it ad hoc} as it does not result from a known particle physics model, however it does results potential which gives the right behaviour for the expanding universe. Also the potential, we obtain, 
 has earlier been considered by different authors for quintessence fields.. In the
context of recent SN1a observation we have shown in this paper that this may
really be a good choice. It will be worthwhile to study the
extrapolation of this model to radiation dominant era to check its consistency
with the big bang nucleosynthesis. Also one should check the consistency of
this model with recent CMB observation. These issues will be addressed later.

\noindent{\bf Acknowledgments}: We would like to thank Claudio Rubano, Bob Jantzen and Diego Pavon  for their valuable comments and suggestions and also for giving references of some important papers. We are also thankful to the anonymous referee for his comments which helped us to improve the clarity of the paper.


\begin{references}
\bibitem{super1}S.Perlmutter, M.Della Valle et al., {\it Nature}, {\bf 391},
 (1998); S.Perlmutter, G.Aldering, G.Goldhaber et al., {\it Astroph. J.},
 {\bf 517} (1999)
\bibitem{super2}P.M. Garnavich, R.P. Kirshner, P. Challis et al., {\it Astroph
. J.}, {\bf 493}, (1998); A.G. Riess, A.V. Philipenko, P. Challis et al.,
{\it Astron. J.}, {\bf 116}, 1009 (1998)
\bibitem{dust}P.S.Drell, T.J.Loredo and I.Wasserman Astrophys.J. {\bf 530},
  593 (2000);
A.G.Riess PASP {\bf 112}, 1284, (2000).
\bibitem{newriess} A.G.Riess, astro-ph/0104455.
\bibitem{cmb}P. Bernadis et al., {\it Nature}, {\bf 404}, 955 (2000);
S.Hanany et al., {\it astro-ph}/0005123; A. Balbi et al., 
{\it astro-ph}/0005124.
\bibitem{cal}R.R.Caldwell, R.dave and P.J.Steinhardt,
{\it{Phys. Rev.
Lett}}, {\bf 80}, 1582 (1998)
\bibitem{peebetc}P.J.E.Peebles and B.Ratra,
{\it{Astrophys.J.Lett.}}, {\bf
325}, L17, (1988); P.G.Ferreira and M.Joyce,
{\it{Phys.Rev.Lett.}}, {\bf
79}, 4740 (1987); E.J.Copeland, A.R.Liddle and
D.Wands,
{\it{Phys.Rev.D}}, {\bf 57}, 4686 (1988).
\bibitem{stei}P.J. Steinhardt, L.Wang and I.Zlatev,
{\it{Phys.Rev.Lett.}}, {\bf 59},
123504 (1999).
\bibitem{zla}I.Zlatev, L.Wang and P.J.Steinhardt,
{\it{Phys.Rev.Lett.}},
{\bf 82}, 896 (1999).
\bibitem{wett}C.Wetterich, Nucl.Phys.B {\bf 302}, 668, (1988).
\bibitem{joyce1}P.G.Ferreira and M.Joyce, Phys.Rev.Lett {\bf 79}, 4740,(1987);
Phys.Rev.D, {\bf 58}, 023503 (1998).
\bibitem{ratra}B.Ratra and P.J.E.Peebles Phys.Rev.D, {\bf 37}, 3406 (1988).
\bibitem{barr}T.Barreiro, E.J.Copeland and N.J.Nunes, Phys.Rev.D, {\bf 61},
127301 (2000). 
\bibitem{johri}Vinod B. Johri, astro-ph/0108247.
\bibitem{sahni1}T.D.Saini, S.Raychaudhuri, V.Sahni and A.A.Starobinsky,
Phys.Rev.Lett., {\bf 85}, 1162 (2000).
\bibitem{sahni2}V.Sahni and L.Wang, Phys.Rev.D, {\bf 59}, 103517 (2000).
\bibitem{lopez}L.Lopez and T.Matos, Phys.Rev.D {\bf 62}, 081302 (2000).
\bibitem{bento1}M.C. Bento, O. Bertolami, N.C. Santos astro-ph/0106405.
\bibitem{sen1}A.A.Sen, Indrajit Chakrabarty and T.R.Seshadri, gr-qc/0005104,
To appear in Gen.Rel.Grav.
\bibitem{rubano}Claudio Rubano, Paolo Scudellaro astro-ph/0103335.
\bibitem{axion}J.E.Kim JHEP {\bf 9905}, 022 (1999).
\bibitem{bertol}N.Bertolo and M.Pietroni, Phys.Rev.D, {\bf 61}, 023518 (1999).
\bibitem{bertolami}O.Bertolami and P.J.Martins, Phys.Rev.D, {\bf 61}, 064007
  (2000).
\bibitem{ritis}R.Ritis, A.A.Marino, C.Rubano and P.Scudellaro, Phys.Rev.D,
  {\bf 62}, 043506 (2000).
\bibitem{uzan}J.P.Uzan Phys.Rev.D, {\bf 59}, 123510 (1999).
\bibitem{ban}N.Banerjee and D.Pavon, Phys.Rev.D, {\bf 63}, 043504 (2001);
Class.Quant.Grav, {\bf 18}, 593 (2001).
\bibitem{amend1}L.Amendola, Phys.Rev.D, {\bf 62}, 043511 (2000).
\bibitem{amend2}L.Amendola, Phys.Rev.D, {\bf 60}, 043501 (1999).
\bibitem{bacci}F.Perrotta, C.Baccigalupi and S.Matarrese, Phys.Rev.D, {\bf
    61}, 023507 (2000).
\bibitem{chiba}T.Chiba, Phys.Rev.D, {\bf 60}, 083508 (1999).
\bibitem{bacci2}C.Baccigalupi, S.Matarrese and F.Perrotta, Phys.Rev.D, {\bf
    62}, 123510 (2000).
\bibitem{gasper1}M.Gasperini, gr-qc/0105082.
\bibitem{gasper2}M.Gasperini, F.Piazza and G.Veneziano, gr-qc/0108016.
\bibitem{riaz}A.Riazuelo and J.Uzan, astro-ph/0107386.
\bibitem{sen2}A.A.Sen, S.Sen and S.Sethi, Phys.Rev.D, {\bf 63}, 107501 (2001).
\bibitem{sen3}S.Sen and A.A.Sen, Phys.Rev.D, {\bf 63}, 124006 (2001).
\bibitem{sen4}A.A.Sen and S.Sen, Mod.Phys.Lett.A, {\bf 16}, 1303, (2001).
\bibitem{dil}P.Binetruy Phys.Rev.D {\bf 60}, 063502 (1999).
\bibitem{eym}M.C.Bento and O.Bertolami Gen.Rel.Grav. {\bf 31}, 1461 (1999).
\bibitem{zipav}W.Zimdahl and D.Pavon, astro-ph/0105479.
\bibitem{tochini}D.Tocchini-Valentini and L.Amendola, astro-ph/0108143.
\bibitem{turriess}M.S.Turner and A.G.Riess astro-ph/0106051.
\bibitem{ellis}G.F.R Ellis and M.Madsen Class.Quant.Grav., {\bf 8}, 667 (1991).\bibitem{uggla}C.Uggla, R.T.Jantzen abd K.Rosquist Gen.Rel.Grav. {\bf 25}, 409 (1993). 
\bibitem{bacci2}C.Baccigalupi, A.Balbi, S.Matarrese, F.Perrotta and N.Vittorio astro-ph/0109097.
\bibitem{eqn}G.Efstathiou MNRAS {\bf 342}, 810 (2000);
L.Amendola Phys.Rev.Lett {\bf 86}, 196 (2001);
R.Bean and A.Melchiorri astro-ph/0110472.
\bibitem{copecross}P.S.Corasaniti and E.J.Copeland astro-ph/0107378.
\bibitem{hamuy}M.Hamuy et.al, Astrophys.J, {\bf 109}, 1 (1995); 
M.Hamuy et.al, Astrophys.J., {\bf 112}, 2391 (1996).
\end{references}
\end{document}